\documentclass[12pt]{article}
\usepackage{geometry,graphicx} 
\usepackage{amssymb,amsmath,amsthm,mathtools,url}
\newcommand{\Prob} {\mbox{$\rm{Prob}$\,}}

\title{Rate Matrix Estimation From Site Frequency Data}
\author{Conrad J. Burden$^{1, 2}$ and Yurong Tang$^1$\\
$^1$ Mathematical Sciences Institute, Australian National University \\
$^2$ Research School of Biology, Australian National University}

\begin{document}

\maketitle
\abstract{A procedure is described for estimating evolutionary rate matrices from observed site frequency data.  The procedure assumes 
(1) that the data are obtained from a constant size population evolving according to a stationary Wright-Fisher model;  
(2) that the data consist of a multiple alignment of a moderate number of sequenced genomes drawn randomly from the population; and 
(3) that within the genome a large number of independent, neutral sites evolving with with a common mutation rate matrix can be identified.  
No restrictions are imposed on the scaled rate matrix other than that the off-diagonal elements 
are positive and $<<1$, and that the rows sum to zero.  In particular the rate matrix is not assumed to be reversible.  The key to the method is an approximate stationary 
solution to the forward Kolmogorov equation for the multi-allele neutral Wright-Fisher model in the limit of low mutation rates.  
}


\section{Introduction}
\label{sec:Introduction}

This paper is a continuation of previous work~\cite{Burden:2016fk} in which an approximate solution to the forward Kolmogorov equation to the multi-allelic neutral Wright-Fisher 
model is derived for the biologically relevant case of low mutation rates.  Herein we address the problem of estimating a mutation rate matrix from site frequency data. 
The data is assumed to take the form of a multiple alignment of independent, neutrally evolving genomic sites sequenced from a moderate number of individuals 
chosen independently from a large effective population.  

For an alphabet of size $K$ alleles the general mutation rate matrix $Q$ has $K(K - 1)$ free parameters, which equates to 12 free parameters for the genomic 
alphabet $\{A, C, G, T\}$.  Classical estimates of mutation rates~\cite{watterson1975number,ewens1974note}, and more recent treatments of the problem 
(see ~\cite{roychoudhury2010sufficiency} and references therein) have been concerned primarily with estimating an overall mutation rate, 
generally denoted by $\theta$, whereas the current paper aims to estimate all parameters of the rate matrix $Q$.  
The equivalent estimation problem for $K= 2$ alleles has been solved by Vogl~\cite{vogl2014estimating} for neutral sites and 
Vogl and Bergman~\cite{vogl2015inference} when selection is included.  

A $2 \times 2$ rate matrix has a total of 2 free parameters to estimate and is necessarily reversible, which simplifies the problem considerably.  The innovation which 
allows us to deal with the $K > 2$ cases is an interpretation of the non-reversible part of the rate matrix as a set of fluxes of probability around closed paths 
in the solution-space simplex of the forward Kolmogorov equation~\cite{Burden:2016fk}.  Section~\ref{sec:paramQ} sets out a convention for parameterising 
the general $K \times K$ mutation rate matrix $Q$ which exploits this interpretation.  When $K = 4$, for instance, we arrive at 3 independent probabilities defining 
the stationary Markov state, 6 parameters specifying 
the remaining degrees of freedom in the reversible part of $Q$, and 3 probability fluxes specifying the non-reversible part, which sums to the required 12 parameters.  
Section~\ref{sec:approxWFsoln} summarises 
our previously reported approximate stationary solution to the forward Kolmogorov for the multi-alleleic neutral-evolution Wright-Fisher model~\cite{Burden:2016fk}.  
Because only low mutation rates are considered the solution can be specified as a set of line densities on the edges and point masses at the corners of the 
$(K - 1)$-dimensional simplex over which the stationary distribution is defined.  

The procedure for estimating the parameters of $Q$ from site frequency data is described in Section~\ref{sec:parameterEstimation}.  Maximum likelihood estimates 
are obtained assuming the data to consist of counts of allele frequencies observed in a finite sample of individuals assumed to be chosen at random from the population.  
Interestingly, RoyChoudhury and Wakeley~\cite{roychoudhury2010sufficiency} come close to providing the equivalent estimate for the restricted case of a 
parent-indpendent rate matrix, but only specify the overall scale $\theta$ and not the complete rate matrix, which, for their restricted case, has 
$K$ parameters and is reversible.  Our estimates are tested using synthetic data for $K = 3$ and $K = 4$ rate matrices in Section~\ref{sec:Results}.  
Conclusions are summarised in Section~\ref{sec:Conclusions}.  


\section{Parameterisation of the rate matrix $Q$}
\label{sec:paramQ}

Suppose we are given any $K \times K$ rate matrix $Q$ whose elements $Q_{ab}$, where $a, b = 1, \ldots K$, must satisfy
\begin{equation}
Q_{ab} \ge 0, \quad \mbox{for $a \ne b$, and} \quad \sum_{b = 1}^K Q_{ab} = 0.    
\end{equation}
These constraints imply that $K(K - 1)$ parameters are necessary to specify $Q$.  Inspired by the results of~\cite{Burden:2016fk} we begin our analysis by 
constructing a parameterisation consistent with the decomposition of $Q$ into a reversible part \cite{lanave1984new,tavare1986some} 
and a flux part, that is,  
\begin{equation}
Q = Q^{\rm GTR} + Q^{\rm flux}. 		\label{QGTRplusFlux}
\end{equation}
The flux part represents a set of fluxes of probability around closed paths between subsets of 3 alleles once the Markovian process has settled into its stationary state.   

Let us assume that $Q$ has a unique stationary state $\pi^{\rm T} = (\pi_1 \ldots \pi_K)$ satisfying 
\begin{equation}
\pi_a \ge 0, \quad \sum_{a = 1}^K \pi_a = 1, \quad \sum_{a = 1}^K \pi_a Q_{ab} = \pi_b.    \label{PiProperties}
\end{equation}
A necessary condition for a unique $\pi^{\rm T}$ to exist is that $Q_{ab} > 0$ for all $a \ne b$.  One would expect this to include any biologically realistic model.  
For an evolving population in its stationary state, the rate of mutations from allele-$a$ to allele-$b$ at any genomic site is $\pi_a Q_{ab}$.  

Define parameters $C_{ab}$ and $\Phi_{ab}$ by  
\begin{equation}
C_{ab} = \pi_a Q_{ab} + \pi_b Q_{ba}, \qquad \Phi_{ab} = \pi_a Q_{ab} - \pi_b Q_{ba}.  \label{CAndPhiDef}
\end{equation}
It is easy to check that 
\begin{equation}
Q_{ab} = \tfrac{1}{2} (C_{ab} + \Phi_{ab})/\pi_a. 			\label{Qparameterisation}
\end{equation}
Hence $Q$ can be decomposed according to Eq.~(\ref{QGTRplusFlux}) where 
\begin{equation}
Q_{ab}^{\rm GTR} = \tfrac{1}{2} C_{ab}/\pi_a, 			\label{QGTRparameterisation}
\end{equation}
satsifies the time-reversible condition 
$\pi_a Q_{ab}^{\rm GTR} = \pi_b Q_{ba}^{\rm GTR}$, 
and 
\begin{equation}
Q_{ab}^{\rm flux} = \tfrac{1}{2} \Phi_{ab}/\pi_a. 			\label{Qfluxparameterisation} 
\end{equation}  
It is clear from Eq.~(\ref{CAndPhiDef}) that $\Phi_{ab}$ is the net flux of probability per unit time 
from allele-$a$ to allele-$b$.  

\begin{figure}[t]
   \centering
   \includegraphics[width=\textwidth]{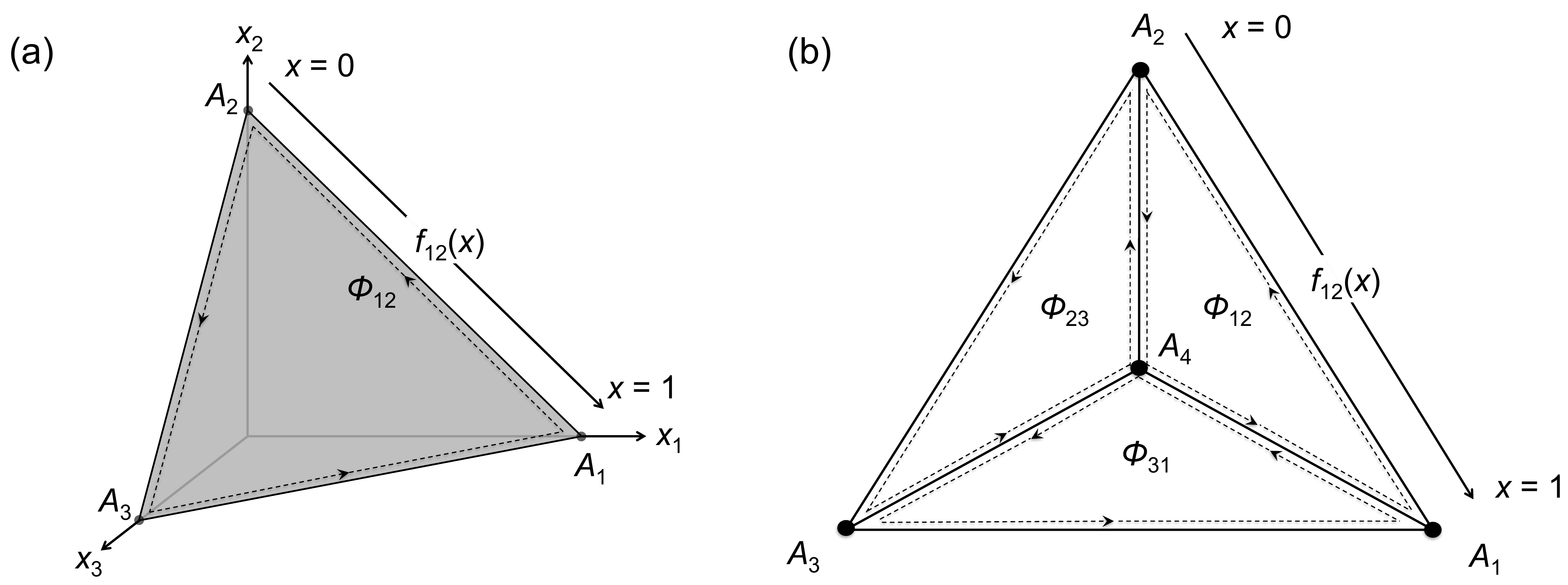} 
   \caption{The simplex on which the solution to the forward Kolmogorov equation for the multi-allele Wright-Fisher model is defined 
   for (a) $K = 3$ alleles and (b) $K = 4$ alleles.  The corners labelled $A_1$, $A_2$, etc.\ indicate the co-ordinates corresponding to non-segregating 
   sites at which  the allele specified is prevalent throughout the population.  The probability fluxes $\Phi_{ab}$ and line densities $f_{ab}(x)$ are 
   explained in the text.}
   \label{fig:Simplices}
\end{figure}

Note that there are certain dependencies between the parameters $\pi_a$, $C_{ab}$ and $\Phi_{ab}$.  Firstly, the normalisation in Eq.~(\ref{PiProperties}) 
implies that only $K - 1$ components of $\pi_a$ are independent, i.e. 
\begin{equation}
\pi_K = 1 - \sum_{i = 1}^{K - 1} \pi_i.		\label{PiK}
\end{equation}
Secondly, $C_{ab} = C_{ba}$, and it follows from the properties of $Q$ that $\sum_{b = 1}^K C_{ab} = 0$.  Thus $C_{ab}$ is a symmetric matrix whose 
diagonal elements are given in terms of its off-diagonal elements via
\begin{equation}
C_{aa} = - \sum_{b \ne a}  C_{ab},\qquad a, b =1, \ldots, K.	\label{Caa}
\end{equation}
Thirdly, $\Phi_{ab} = -\Phi_{ba}$, and it follows from the properties of $Q$ 
that $\sum_{b = 1}^K \Phi_{ab} = 0$.  Thus $\Phi_{ab}$ is an antisymmetric matrix whose rows sum to zero, that is, the final row and column of 
$\Phi_{ab}$ are given in terms of the remaining elements via 
\begin{equation}
\Phi_{iK} = - \Phi_{Ki} = - \sum_{j \ne i} \Phi_{ij},\qquad i, j = 1, \ldots, K - 1.	\label{PhiKi}
\end{equation}
Equation~(\ref{PhiKi}) is at statement that, in the steady state, the net flux of probability from any allele is zero.  For $K = 3$ alleles there is only 
one independent flux, $\Phi_{12}$, and the elements of $Q$ are 
\begin{equation}
Q = \frac{1}{2}\left( \begin{array}{ccc}
\dfrac{-\,C_{12} - C_{13}}{\pi_1}				&	\dfrac{C_{12} + \Phi_{12}}{\pi_1}			&	\dfrac{C_{13} - \Phi_{12}}{\pi_1}	\\ \\
\dfrac{C_{12} - \Phi_{12}}{\pi_2}			&	\dfrac{-\,C_{12} - C_{23}}{\pi_2}				&	\dfrac{C_{23} + \Phi_{12}}{\pi_2}	\\ \\
\dfrac{C_{13} + \Phi_{12}}{1 - \pi_1 - \pi_2}	&	\dfrac{C_{23} - \Phi_{12}}{1 - \pi_1 - \pi_2}	&	\dfrac{-\,C_{13} - C_{23}}{1 - \pi_1 - \pi_2}		
\end{array} \right).		\label{K3Parameters}
\end{equation}
For $K = 4$ alleles here are three independent fluxes  $\Phi_{12}$, $\Phi_{23}$ and $\Phi_{31}$ as illustrated in Fig.~\ref{fig:Simplices}.

To summarise, the general rate matrix $Q$ can be parameterised via Eqs.~(\ref{QGTRplusFlux}), (\ref{QGTRparameterisation}) and 
(\ref{Qfluxparameterisation}) using the following minimal set of parameters: 
\begin{equation}
\begin{array}{lrr}
\pi_i, & i = 1, \ldots, K - 1: & K - 1\mbox{ parameters}; \\
C_{ab} = C_{ba}, & 1 \le a < b \le K: & \tfrac{1}{2}K(K - 1) \mbox{ parameters}; \\
\Phi_{ij} = -\Phi_{ji}, & 1 \le i < j \le K - 1: & \tfrac{1}{2}(K - 1)(K - 2) \mbox{ parameters},		\label{parameterList}
\end{array}
\end{equation}
with the remaining, unspecified parameters given by Eqs.~(\ref{PiK}), (\ref{Caa}) and (\ref{PhiKi}).  The total number of independent parameters 
listed in Eq.~(\ref{parameterList}) is $K(K - 1)$, as required.  The requirement that the off-diagonal elements of $Q$ be positive implies the further constraints 
on the parameter space that 
\begin{equation}
\pi_a \ge 0, \quad C_{ab} \ge 0, \quad \left| \Phi_{ab} \right| \le C_{ab}, \qquad  1 \le a < b \le K.  \label{paramConstr}
\end{equation}

The remainder of this paper is concerned with estimating the $K(K - 1)$ parameters of a genomic evolutionary rate matrix from site frequency data assuming a 
population whose genome includes a large number of independent sites that have evolved to stationarity according to a neutral evolution Wright-Fisher model.  


\section{Approximate solution to the neutral multi-allele Wright-Fisher model}
\label{sec:approxWFsoln}

We consider the neutral evolution Wright-Fisher model for $K$ alleles, labelled $A_1$ \ldots $A_K$ (see, for example, Section~4.1 of ref.~\cite{etheridge2011some}).  
Given a haploid population of size $N$ (or monoecious diploid population of size $N/2$), 
let the number of individuals of type $A_a$ at time step $\tau$ be $Z_a(\tau)$ for discrete times $\tau = 0, 1, 2, \ldots$.  
Also, let $u_{ab}$ be the probability of an individual making a transition from $A_a$ to $A_b$ in a single time step, where $u_{ab} \ge 0$ and 
$\sum_{b = 1}^K u_{ab} = 1$.  Writing $\mathbf Z(\tau) = (Z_1(\tau), \ldots Z_K(\tau))$, the multi-allele neutral Wright-Fisher model is defined by the transition 
matrix from an allele frequency $\mathbf i = (i_1 , \ldots, i_K)$ to an allele frequency $\mathbf j = (j_1 , \ldots, j_K)$ in the population given by  
\begin{equation}
\Prob(\mathbf Z (\tau + 1) = \mathbf j | \mathbf Z (\tau) = \mathbf i) = 
            \frac{N!}{\prod_{a = 1}^K j_a!} \prod_{a=1}^K{\psi(\mathbf{i}, a})^{j_a}  ,  \label{eq:FullPij}
\end{equation}
where $\sum_{a = 1}^K i_a = \sum_{a = 1}^K j_a = N$, and 
\begin{equation}
\psi(\mathbf{i}, a) = \frac{i_a}{N}  \left(1 - \sum_{b \ne a} u_{ab} \right) + \sum_{b \ne a} \frac{i_b}{N} u_{ba} = \sum_{b = 1}^K \frac{i_b}{N} u_{ba}.  \label{psiDef}
\end{equation}

The usual diffusion limit is obtained by defining random variables $X_a(t) = Z_a(\tau)/N$ equal to the relative proportion of type-$A_a$ 
alleles within the population at continuous time $t = \tau/N$.  The limit $N \rightarrow \infty$ and $u_{ab} \rightarrow 0$ for $a \ne b$ is taken in such a 
way that the $K \times K$ instantaneous rate matrix $Q$, whose elements are defined by 
\begin{equation}
Q_{ab} = N (u_{ab} - \delta_{ab}),	\label{qToU}
\end{equation}
remains finite.  This limit leads to a forward Kolmogorov equation for the density function $f_{\bf X}(x_1, \ldots, x_{K-1}; t)$ of the vector of continuous 
random variables $X_1(t), \ldots, X_{K - 1}(t)$.  
The function $f_{\bf X}$ is is defined over the simplex (see Fig.~\ref{fig:Simplices}) 
\begin{equation}
{\cal S} = \left\{ (x_1, \ldots, x_K) : x_1, \ldots, x_K \ge 0, \, \sum_{a = 1}^K x_a = 1\right\}.  \label{simplex}
\end{equation} 
Further details of the equation are summarised in~\cite{Burden:2016fk}.  

Solution of the forward Kolmogorov equation for an arbitrary rate matrix $Q$ and $K \ge 3$ alleles, even for the stationary distribution when 
$\partial f_{\bf X}/\partial t$ is set to zero, remains an unsolved problem.  However, 
in \cite{Burden:2016fk} we derive an approximate stationary distribution in the biologically realistic limit of slow but otherwise arbitrary mutation rates, 
that is for $0 \le Q_{a b} <<1$ for $a \ne b$.  Our analysis is based on the result that, in the limit $Q_{ab} \rightarrow 0$, the stationary probability 
distribution is concentrated close to the edges of the simplex $\cal S$, and can therefore be represented accurately as a set of line 
densities defined along those edges.   Suppose we label the corners of $\cal S$ by the allele prevalent within the population at that corner, so the corner 
$A_1$ corresponds to to the co-ordinate $(x_1, \ldots, x_K) = (1, 0, \ldots, 0)$, and so on, as illustrated in Fig.~\ref{fig:Simplices}.   On the edge joining 
corner $A_a$ to corner $A_b$ define a line density $f_{ab}(x)$ for each pair of indices $a$ and $b$.
We will adopt the convention that the argument $x$ is the relative proportion of type-$a$ alleles, and $1 - x$ is the relative proportion of type-$b$ alleles.  
The relative proportion of the remaining $K- 2$ alleles along this edge is zero.  

The line densities are given in terms of the parameterisation introduced in Section~\ref{sec:paramQ} as (see Eq.~(53) of \cite{Burden:2016fk}) 
\begin{equation}
f_{ab}(x) = C_{ab}\left(\frac{1}{x} + \frac{1}{1 - x}\right) - \Phi_{ab}\left(\frac{1}{x} - \frac{1}{1 - x}\right).		\label{fabSolution}
\end{equation}
Note that $f_{ab}(x) = f_{ba}(1 - x)$.  A necessary condition for the line density to be an accurate representation of the exact solution is that 
\begin{equation}
Q_{ab} \times \left| \log(x) + \log(1 - x) \right| << 1.  
\end{equation}
Thus the approximation loses accuracy as $x$ approaches 0 or 1, that is, in the vicinity of the corners of $\cal S$.  
However, for a population of size $N$, the value of the stationary distribution 
at the corners of the simplex corresponding to the discrete problem defined by Eq.~(\ref{eq:FullPij}) is, to a good approximation,  (see Eq.~(56) of \cite{Burden:2016fk}) 
\begin{equation}
P(A_a) = \Prob(Z_a = N, Z_b = 0 \mbox{ for } b \ne a) = \pi_a - \sum_{b \ne a} C_{ab} \log N.  \label{PASolution}
\end{equation}
As a rule of thumb, we have observed in numerical simulations that Eqs.~(\ref{fabSolution}) and (\ref{PASolution}) provide a very good approximation 
to the stationary state of the discrete model provided the off-diagonal elements of $Q$ are less than $10^{-2}$.  The approximate solution is normalised 
in the sense that 
\begin{equation}
\lim_{N \rightarrow \infty}\left( \sum_{1 \le a < b \le K} \int_{1/N}^{1 - 1/N} f_{ab}(x)\, dx + \sum_{a = 1}^K P(A_a) \right) = 1.  
\end{equation} 


\section{Parameter Estimation}
\label{sec:parameterEstimation}

Assume we have a data set in the form of a site frequency spectrum (SFS) obtained by sampling $L$ independent neutrally evolving sites within 
the genomes of $M$ individuals from a population of size $N >> M$.  Typically $L$ might be at least $10^3$, $M$ in the range 10 to 100, and $N$ is 
ideally essentially infinite in the sense that the diffusion limit forward-Kolmogorov equation is appropriate.  $L$, $M$ and $N$ are known fixed parameters.  
At each genomic site $l$, define a vector of non-negative integer valued random variables 
\begin{equation}
\mathbf{Y}^{(l)} = Y_1^{(l)}, \ldots, Y_K^{(l)}, \qquad l = 1, \ldots L, 
\end{equation}
where $Y_a^{(l)}$ equal to the number of times allele type-$a$ occurs within the sampled individuals.  Clearly $\sum_{a = 1}^K Y_a^{(l)} = M$, so the 
data at any given genomic site can be specified as a point in a $(K - 1)$-dimensional simplex lattice.  

Given an observed data set $(\mathbf{y}^{(1)}, \ldots, \mathbf{y}^{(L)})$, the log-likelihood is 
\begin{equation}
{\cal L}(\pi, C, \Phi | \mathbf{y}^{(1)}, \ldots , \mathbf{y}^{(L)}) = \sum_{l = 1}^L \log \Prob(\mathbf{Y}^{(l)} = \mathbf{y}^{(l)} \, | \, \pi, C, \Phi), 
\end{equation}
where the triplet $(\pi, C, \Phi)$ represents the $K(K - 1)$ parameters of Eq.~(\ref{parameterList}).  The probabilities occurring in this 
sum are calculated under the assumption that the data is sampled randomly from a population with genomic sites distributed according to the 
approximate stationary solution of  Section~\ref{sec:approxWFsoln}.  Below we show that these probabilities are given by
\begin{eqnarray}
\lefteqn{\Prob(\mathbf{Y}^{(l)} = \mathbf{y} \, | \, \pi, C, \Phi) = } \nonumber \\
&&\begin{cases} 
\pi_a - H_M \sum_{b\ne a}C_{ab}, & \mbox{if $y_a = M$ and all other} \\
						& \quad \mbox{components of $\mathbf{y}$ are zero,} \\  \\
C_{ab} \left( \frac{1}{y} + \frac{1}{M - y} \right) - \Phi_{ab} \left( \frac{1}{y} - \frac{1}{M - y} \right) & \mbox{if $y_a = y$, $y_b = M - y$ and all} \\
						& \quad \mbox{other components of $\mathbf{y}$ are zero,}  \\ \\
0  & \mbox{otherwise},
\end{cases}									\label{probYsGivenParameters}
\end{eqnarray}	
where 
\begin{equation}
H_M = \sum_{y = 1}^{M - 1} \frac{1}{y}. 
\end{equation}
This distribution generalises the corresponding  $K = 2$ distribution found previously by Vogl, namely Eq.~(13) of \cite{vogl2014estimating}, for which   
Vogl and Bergman~\cite{vogl2015inference} propose the name ``generalised RoyChoudhury-Wakeley distribution".  
There is no $\Phi_{ab}$ contribution for $K = 2$ as any $2 \times 2$ rate matrix is automatically reversible.  
Furthermore RoyChoudhury and Wakeley's distribution, namely Eq.~(10) of \cite{roychoudhury2010sufficiency}, is actually the restriction of 
the second part of Eq.~(\ref{probYsGivenParameters}) to the case of a $K$-allele parent-independent 
rate matrix, $Q_{ab} = \frac{1}{2}\theta \pi_b$, for arbitrary $K \ge 2$ and a canonical scaling parameter $\theta << 1$.  In terms of our parameterisation 
Eq.~(\ref{parameterList}) this corresponds to setting $C_{ab} = \theta \pi_a \pi_b$ and $\Phi_{ij} = 0$.   
The parent-independent rate matrix is also reversible, and includes the most general $2 \times 2$ rate matrix when $K = 2$.  
We therefore propose that the name generalised RoyChoudhury-Wakeley-Vogl-Bergman distribution could be appropriately applied to 
Eq.~(\ref{probYsGivenParameters}).  
A proof of Eq.~(\ref{probYsGivenParameters}) follows.  

\begin{proof}
Consider the three cases in turn.

\noindent {\bf Case I:} Sites for which exactly 1 component of $\mathbf{y}^{(l)}$ is non-zero.  Suppose site-$l$ is non-segregating with allele $A_a$ occurring 
in all individuals within the sample.  In this case one can reduce the continuum diffusion limit to an effective 2-allele model in which all alleles except $A_a$ 
are combined into a single allele, $A_{\bar a}$.  As described in Eqs.~(A.4) and (A.5) of Appendix~A of~\cite{Burden:2016fk}, the effective 2-allele model has a rate matrix 
\begin{equation}
\tilde{Q} = \left( \begin{array}{cc}
1 - \tilde{Q}_{a\bar{a}} & \tilde{Q}_{a\bar{a}} \\
\tilde{Q}_{\bar{a} a} & 1 - \tilde{Q}_{\bar{a} a}  
\end{array}\right) , 
\end{equation}
with, 
\begin{equation}
\begin{split}
\tilde{Q}_{a\bar{a}} = \frac{\sum_{b \ne a}\pi_a Q_{ab}}{\pi_a} = \frac{\sum_{b\ne a}C_{ab}}{2\pi_a}, \\
\tilde{Q}_{\bar{a}a} = \frac{\sum_{b \ne a}\pi_b Q_{ba}}{1 - \pi_a} = \frac{\sum_{b\ne a}C_{ab}}{2(1 - \pi_a)} 
\end{split}
\end{equation}
where we have used Eq.~(\ref{Qparameterisation}) and the fact that $\Phi_{ab}$ is an anti-symmetric matrix whose rows sum to zero.  In Section~4 of 
Ref.~\cite{vogl2014estimating} Vogl provides an analysis of the 2-allele model in the small mutation rate limit.  Vogl parameterises the $2 \times 2$ rate 
matrix in terms of two parameters, $\vartheta$ and $\alpha$, which are related to our parameters by $\tilde{Q}_{a\bar{a}} = \vartheta/(2\alpha)$, 
$\tilde{Q}_{\bar{a}a} = \vartheta/[2(1 - \alpha)]$, or equivalently, 
\begin{equation}
\vartheta = \sum_{b\ne a}C_{ab}, \qquad \alpha = \pi_a.  \label{VoglsParameters}
\end{equation}
From Eq.~(29) of \cite{vogl2014estimating}, using the above identification we can immediately read off the required probability 
\begin{equation}
\Prob\left\{\mathbf{Y}^{(l)} = (0, \ldots, Y_a^{(l)} = M, \ldots, 0)\right\} = \pi_a - H_M \sum_{b\ne a}C_{ab}.
\end{equation}

\noindent {\bf Case II:} Sites for which exactly 2 components of $\mathbf{y}^{(l)}$ are non-zero.  Suppose site $l$ is biallelic with alleles $A_a$ and $A_b$ 
occurring $y_a^{(l)} = y$ times and $y_b^{(l)} = M - y$ times respectively within the sample.  Then, from Eq.~(\ref{fabSolution}), 
\begin{eqnarray}
\lefteqn{\Prob\left\{\mathbf{Y}^{(l)} = (0, \ldots, y, \ldots, M - y, \ldots, 0)\right\}} \qquad\nonumber \\
& = & \int_{1/N}^{1 - 1/N} f_{ab}(x) {M\choose y} x^y (1 - x)^{M - y} dx \nonumber \\
& = & {M\choose y}\left[(C_{ab} - \Phi_{ab}) \int_{1/N}^{1 - 1/N} x^{y - 1} (1 - x)^{M - y} dx \right. \nonumber \\ 
& &        \qquad\qquad\qquad\qquad \left.+ (C_{ab} + \Phi_{ab}) \int_{1/N}^{1 - 1/N} x^y (1 - x)^{M - y - 1} dx \right] \nonumber \\ 
& = & {M\choose y}\left[(C_{ab} - \Phi_{ab}) B(y, M - y + 1) \right. \nonumber \\ 
& &        \qquad\qquad\qquad\qquad \left.+ (C_{ab} + \Phi_{ab}) B(y + 1, M - y) \right]  + O\left(\frac{1}{N}\right), 
\end{eqnarray}
where $B(m, n) = \Gamma(m)\Gamma(n)/\Gamma(m + n)$ is the beta function.  For positive integer arguments $\Gamma(n) = (n - 1)!$, 
which reduces the last line to 
\begin{eqnarray}
\lefteqn{\Prob\left\{\mathbf{Y}^{(l)} = (0, \ldots, y, \ldots, M - y, \ldots, 0)\right\}} \qquad\qquad\qquad\nonumber \\
& \approx & C_{ab} \left( \frac{1}{y} + \frac{1}{M - y} \right) - \Phi_{ab} \left( \frac{1}{y} - \frac{1}{M - y} \right), 
\end{eqnarray}
up to order $1/N$.   

\noindent {\bf Case III:} Sites for which 3 or more components of $\mathbf{y}^{(l)}$ are non-zero.  The assumption that the solution to the 
forward Kolmogorov equation can be represented by a set of line densities on the edges plus point masses at the vertices implies that, 
in the entire the population, no more than two alleles can be represented at any given genomic site.  In other words, the assumed model 
entails that this case will occur with probability zero.  
\end{proof}

Regarding Case III, since 3-allelic and 4-allelic sites are rare in genomes~\cite{cao2015analysis,phillips2015tetra}, we argue that removing 
such sites from the data will not do serious damage to a maximum likelihood estimator of the parameters.  Alternatively, one could reassign 
such data points to the nearest point on an edge of the simplex lattice.  

Now define the following variables: 
\begin{equation}
\begin{split}
L_a &= \sum_{i = 1}^L I\left(Y_a^{(l)} = M, Y_b^{(l)} = 0 \mbox{ for $b \ne a$}\right), \quad a = 1, \ldots , K, \\
L_{ab}(y) &= \sum_{i = 1}^L I\left(Y_a^{(l)} = y, Y_b^{(l)} = M - y, Y_c^{(l)} = 0 \mbox{ for $c \ne a, b$}\right), \\
		&	\qquad\qquad\qquad\qquad\qquad\qquad 1 \le a < b \le K; \, y = 1, \ldots, M - 1,  \\
L_{ab} &= \sum_{y = 1}^{M - 1} L_{ab}(y), 
\end{split}								\label{LDefs}
\end{equation}
where $I(\cdot)$ is the indicator random variable for the event specified.  That is, $L_a$ is a count of the number of non-segregating sites of allele 
type $A_a$, $L_{ab}(y)$ is a count of the number of biallelic polymorphisms with $y$ occurrences of allele $A_a$ and $M - y$ occurrences of 
allele $A_b$, and $L_{ab}$ is the total number of biallelic polymorphisms of type $A_a$-$A_b$.  We will assume the data is such that all sites observed 
are either non-segregating or biallelic, i.e., $\sum_a L_a + \sum_{a<b} L_{ab} = L$.  Then 
\begin{eqnarray}
\lefteqn{\Prob(L_a = l_a, L_{ab}(y) = l_{ab}(y) \, | \, \pi, c, \Phi) = } \nonumber \\
& & \frac{L!}{(\prod_{a = 1}^K l_a!) (\prod_{a<b} \prod_{y = 1}^{M - 1} l_{ab}(y)!)}  
	\left[ \prod_{a = 1}^K \left(\pi_a - H_M \sum_{b \ne a} C_{ab} \right)^{l_a} \right] \times \nonumber \\
 & & \qquad \left[ \prod_{a < b} \prod_{y = 1}^{M - 1}  \left(C_{ab}\left\{\frac{1}{y} + \frac{1}{M - y} \right\} - 
                                                                                         \Phi_{ab} \left\{\frac{1}{y} - \frac{1}{M - y} \right\} \right)^{l_{ab}(y)} \right], \nonumber \\	\label{probAllLs}
\end{eqnarray}
and 
\begin{eqnarray}
\lefteqn{\Prob(L_a = l_a, L_{ab} = l_{ab} \, | \, \pi, c, \Phi) = } \nonumber \\
& & \frac{L!}{(\prod_{a = 1}^K l_a!)(\prod_{a<b}  l_{ab}!)}  
	\left[ \prod_{a = 1}^K \left(\pi_a - H_M \sum_{b \ne a} C_{ab} \right)^{l_a} \right]  \left[ \prod_{a < b}  \left(2 H_M C_{ab} \right)^{l_{ab}} \right].  
							\nonumber \\	\label{probSomeLs}
\end{eqnarray}

Since Eq.~(\ref{probSomeLs}) does not depend on $\Phi$, the observations $L_a$ and $L_{ab}$ are sufficient statistics for estimating $\pi_i$ 
and $C_{ab}$~\cite{ewens1974note,roychoudhury2010sufficiency}.  
Moreover, for any index $a = 1, \ldots, K$, one can again define an effective 2-allele model by partitioning the set of alleles into $A_a$ and an effective 
allele $A_{\bar{a}}$ consisting of the remaining $K - 1$ alleles, and then use Vogl's unbiased maximum-likelihood estimator for $\hat\alpha$ 
(see Eq.~(37) of \cite{vogl2014estimating}) together with Eq.~(\ref{VoglsParameters}) to obtain the estimators 
\begin{equation}
\hat\pi_a = \frac{1}{L}\left(L_a + \frac{1}{2}\sum_{b \ne a} L_{ab} \right).  \label{piEstimate}
\end{equation}
Similarly, one can consider a broader set of partitionings of alleles into two exhaustive disjoint subsets and the corresponding effective 2-allele models, 
together with Vogl's unbiased maximum-likelihood estimator for $\hat\vartheta$ (see Eq.~(36) of \cite{vogl2014estimating}) to obtain the estimators 
\begin{equation}
\hat C_{ab} = \frac{L_{ab}}{2LH_M}.  \label{cEstimate}
\end{equation}
It is a straightforward exercise to confirm using Eqs.(\ref{probYsGivenParameters}) and (\ref{LDefs}) that these are unbiased estimators.  

It remains to estimate the flux parameters $\Phi_{ij}$.  In the following we carry out a numerical maximisation of the log-likelihood 
formed from Eq.~(\ref{probAllLs}) by plugging in the above estimates of $\hat C_{ab}$.  Up to an additive constant this gives 
\begin{eqnarray}
\lefteqn{{\cal L}(\Phi_{ij} | l_a, l_{ab}(1), \ldots, l_{ab}(M - 1))} \nonumber \\
& = & \sum_{a < b} \sum_{y = 1}^{M - 1} l_{ab}(y) \log\left(\hat C_{ab}\left\{\frac{1}{y} + \frac{1}{M - y} \right\} - 
                                                                                         \Phi_{ab} \left\{\frac{1}{y} - \frac{1}{M - y} \right\} \right).  \label{phiEstimateLikelihood}
\end{eqnarray} 
The right hand side of this formula is a function of $\frac{1}{2}(K - 1)(K - 2)$ parameters $\Phi_{ij}$ for $1 \le i < j \le K - 1$, with 
Eq.~(\ref{PhiKi}) used to interpret the terms in the sum for which $b = K$.  


\section{Results}
\label{sec:Results}

We have constructed a number of synthetic datasets to test the efficacy of the above theory.  Since the exact solution of the Forward Kolmogorov equation is 
unknown, starting from an assumed scaled rate matrix $Q$ we first generate numerically a stationary site frequency spectrum from the neutral Wright-Fisher model 
for as large a population $N$ as is practicable.  For this step the full transition matrix, Eq.~(\ref{eq:FullPij}), of size ${N + K - 1 \choose K - 1} \times {N + K - 1 \choose K - 1}$ 
is used.  Each dataset is then created corresponding to a multiple alignment at a large number of $L$ independent genomic sites of the genomes of $M$ individuals sampled 
randomly from the population.  Here we have aimed to satisfy the ideal limit $M << N$ within the constraints of the simulation.  Each genomic site is assumed to be the result of 
the Markov process of the full transition matrix evolving to its stationary state.  Details of the sampling process are described in detail below.  
Finally the parameters of $Q$ for each of 1000 such independently generated datasets are estimated using the theory of Section~\ref{sec:parameterEstimation}, and the results 
presented as histograms.  


\subsection{Synthetic Data: $K = 3$ Alleles} 

For $K = 3$ alleles, and for each rate matrix tested, a numerical stationary solution to the neutral Wright-Fisher model with matrix Eq.~(\ref{eq:FullPij})  
was first created for a population size 
$N = 100$.  Three rate matrices were considered.  Each matrix had the same reversible part, $Q^{\rm GTR}$, corresponding to the parameters (see Eq.~(\ref{K3Parameters}))
\begin{equation}
(\pi_1, \pi_2) = (0.5, 0.3), \qquad (C_{12}, C_{23}, C_{13}) = (0.0003, 0.0006, 0.0004). 	\label{truePiAndC}
\end{equation}
For the single flux parameter which determines $Q^{\rm flux}$, namely $\Phi \equiv \Phi_{12}$, three cases were considered: 
\begin{equation}
\Phi = 0, \quad \Phi = 0.0001, \quad \text{and} \quad \Phi=0.0002. 
\end{equation}

For each value of $\Phi$ total of $1000$ synthetic datasets were constructed, each assuming a sample of $M = 10$ individuals sequenced at $L = 10^5$ 
independent genomic sites.  
At each site $l \in \{1, \ldots , L\}$ and within each dataset the stationary distribution was sampled to establish the relative frequencies of the 3 alleles in the population at that site.  
These relative frequencies were used as parameters of a multinomial distribution from which we sampled the observed counts $y_a^{(l)} \in \{0, \ldots , M\}$ 
of the number of times allele $A_a$ was observed at site $l$.   Any genomic site displaying more than 2 alleles in the sample was discarded, though   
generally this amounted to no more than 1 or 2 tri-allelic SNPs observed per dataset.  
Maximum likelihood estimates $\hat\pi_a$, $\hat C_{ab}$ and $\hat\Phi$ of the parameters 
defining $Q$ were obtained for each dataset using the theory developed in Section~\ref{sec:parameterEstimation} (see Eqs.~(\ref{piEstimate}) to 
(\ref{phiEstimateLikelihood})).  

\begin{figure}[t]
   \centering
   \includegraphics[width=\textwidth]{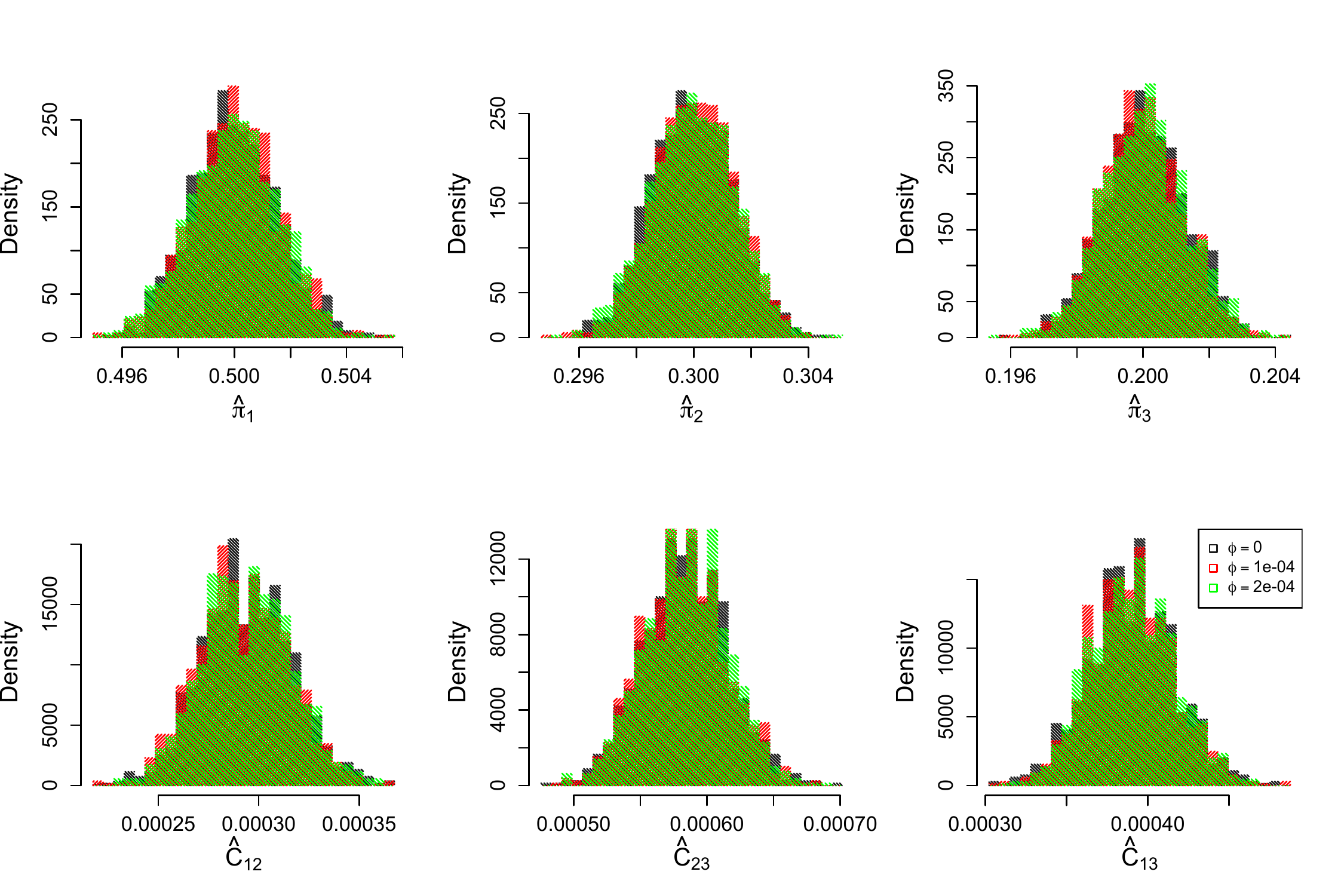} 
   \caption{Histograms of estimates $\hat\pi_a$ and $\hat C_{ab}$ from of the parameters defining the reversible part $Q^{\rm GTR}$ of the rate matrix 
   for $K = 3$ alleles.  True values of the parameters are $(\pi_1, \pi_2, \pi_3) = (0.5, 0.3, 0.2)$ and $(C_{12}, C_{23}, C_{13}) = (0.0003, 0.0006, 0.0004)$.  
   1000 independent datasets were generated for each of 3 values of the flux parameter, $\Phi = 0$, $0.0001$ and $0.0002$.}
   \label{fig:piAndCEstimatePlot}
\end{figure}

Histograms of the estimates $\hat\pi_a$ and $\hat C_{ab}$ are shown in Fig.~\ref{fig:piAndCEstimatePlot}.   The estimates $\hat\pi_a$ are centred about 
their true values, while the estimates $\hat C_{ab}$ tend to be slightly low, perhaps because of the approximate nature of the solution to the forward Kolmogorov 
equation.  The distributions of these parameters are independent of $\Phi$, as expected.  

\begin{figure}[t]
   \centering
   \includegraphics[width=\textwidth]{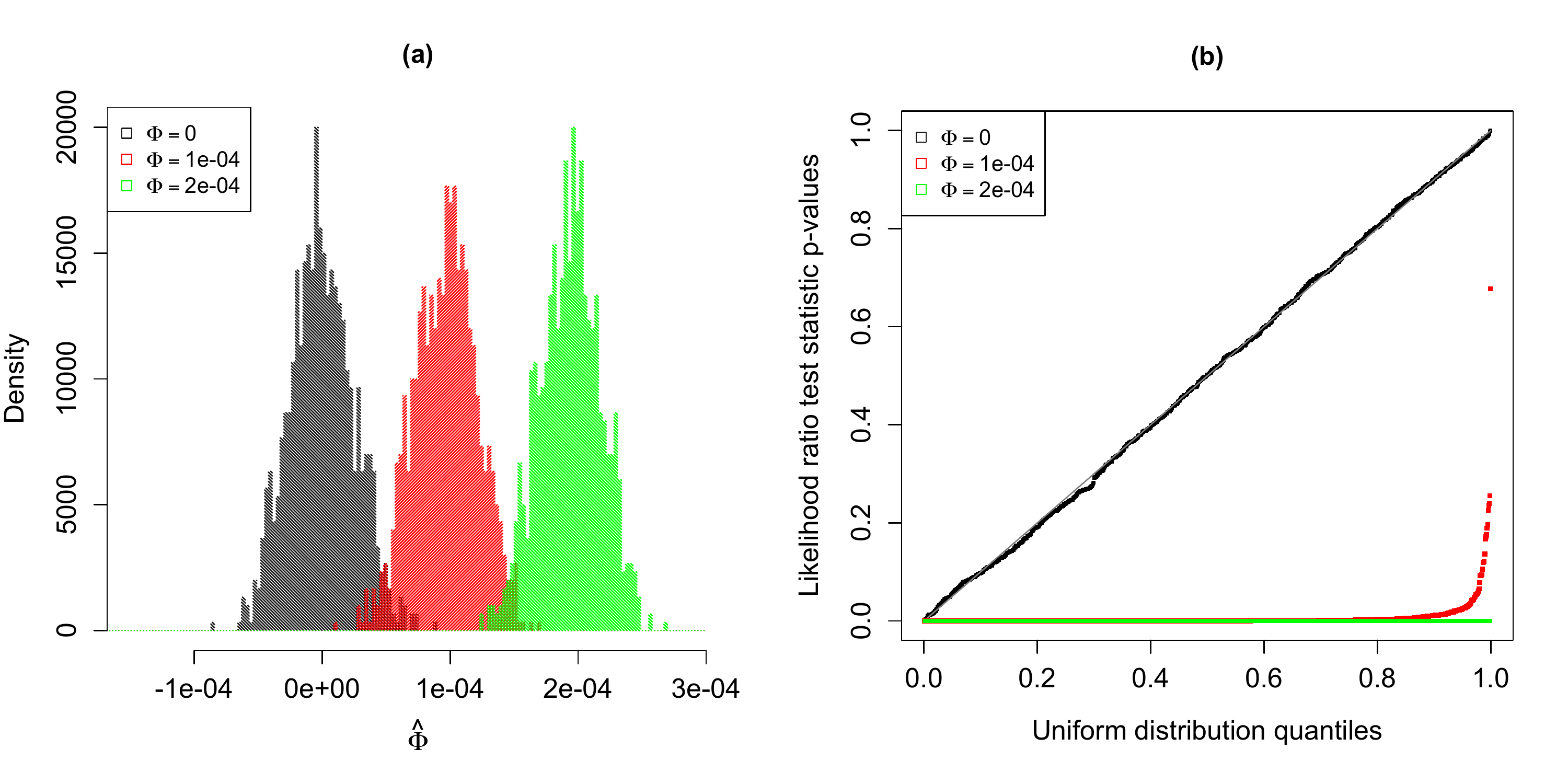} 
   \caption{(a) Histograms of estimates $\hat\Phi$ from of the flux parameter contributing to the non-reversible part $Q^{\rm flux}$ of the rate matrix 
   for $K = 3$ alleles.  True values of the parameters are as in the caption to Fig.~\ref{fig:piAndCEstimatePlot}. (b) Ordered 1-sided p-values calculated from 
   the likelihood ratio test statistic Eq.~(\ref{KEq3LRTStatistic}) plotted against uniform quantiles. For each true value of $\Phi$ p-values are calculated from 
   1000 synthetic datasets. }
   \label{fig:phiEstimatePlot}
\end{figure}

Histograms of the estimates $\hat\Phi$ are shown in Fig.~\ref{fig:phiEstimatePlot}.   These estimates are well centred about their true values.  
Under the null hypothesis that $\Phi = 0$ the likelihood ratio test statistic 
\begin{equation}
-2 \left[{\cal L}(0 | l_a, l_{ab}(1), \ldots, l_{ab}(M - 1)) - {\cal L}(\hat\Phi_{ij} | l_a, l_{ab}(1), \ldots, l_{ab}(M - 1)) \right], \label{KEq3LRTStatistic}
\end{equation} 
where ${\cal L}(\cdot)$ is given by Eq.~(\ref{phiEstimateLikelihood}), 
should have a chi-squared distribution with 1 degree of freedom.  For each true value of $\Phi$ and each dataset a one-sided p-value was calculated 
from this statistic using the upper tail of the chi-squared distribution.  Shown in the right-hand panel of Fig.~\ref{fig:phiEstimatePlot} for each 
$\Phi$-value are ordered p-values from the 1000 datasets, plotted against quantiles of a uniform distribution.  For the $\Phi = 0$ datasets the 
p-values have a uniform distribution as required, while the p-values for the remaining two values of $\Phi$ are suitably small.  


\subsection{Synthetic Data: $K = 4$ Alleles} 

A numerical stationary solution to the neutral Wright-Fisher model with transition matrix Eq.~(\ref{eq:FullPij}) was created for a population size 
$N = 30$ for each of the following 4 rate matrices, defined by the parameters in Table~\ref{tab:RateMatrixParams}: 
\begin{description}
\item[GTR.]  The $4 \times 4$ general time reversible matrix has 9 independent parameters, which can be specified as the $\pi_i$ and $C_{ab}$ defined 
in Eq.~(\ref{parameterList}).  The first 9 parameters in the first column of Table~\ref{tab:RateMatrixParams} are chosen to reproduce very approximately 
the time-reversible matrix in Table~5 of \cite{lanave1984new} estimated from the rat-mouse phylogeny.  The resulting scaled GTR matrix is  
\begin{equation}
Q^{\rm GTR} = \left( \begin{array}{rrrr}
	-5.375 	& 1.875 		& 2.000 	& 1.5000 \\
	2.500 	& -17.500 		& 0.333 	& 14.667 \\
	16.000 	&  2.000 		& -21.000	& 3.000 \\
	2.400 	&  17.600 		&  0.600 	& -20.600
\end{array} \right)    \times 10^{-4}.
\end{equation}
Note that Lavane et al's rate matrix is a per-generation rate and differs from our scaled matrix by a factor of $10^5$, which, by Eq.~(\ref{qToU}), corresponds to 
assuming the effective population size of Lanave et al's data set to be $10^5$.  
\item[GRM.] The most general $4 \times 4$ rate matrix has 12 parameters, which can be specified as the parameters $\pi_i$, $C_{ab}$ and $\Phi_{ij}$ defined in 
Eq.~(\ref{parameterList}).  The general rate matrix $Q^{\rm GRM}$ in our simulations uses the same reversible part as $Q^{\rm GTR}$, plus a non-reversible part using 
the $\Phi_{ij}$ specified in the first column of Table~\ref{tab:RateMatrixParams}, which are chosen arbitrarily subject to the constraint that they should lie within 
the allowable range specified by Eq.~(\ref{paramConstr}).  This gives 
\begin{equation}
Q^{\rm GRM} = \left( \begin{array}{rrrr}
	-5.375 	& 3.125 		& 2.125 	& 0.1250 \\
	0.833 	& -17.500 		& 0.583 	& 16.083 \\
	15.000 	&  0.500 		& -21.000	& 5.500 \\
	4.600 	&  15.900 		&  0.100 	& -20.600
\end{array} \right)    \times 10^{-4}.
\end{equation}
\item[SS.] A strand-symmetric rate matrix is one which is symmetric under simultaneous interchange of nucleotides $A$ with $T$ and $C$ with $G$.   
Strand symmetry imposes certain constraints on the parameters listed in Eq.~(\ref{parameterList}), namely, 
\begin{equation}
\begin{split}
& \pi_C = \pi_G = 0.5 - \pi_A, \\
& C_{AC} = C_{GT}, \qquad C_{AG} = C_{CT}, \\
& \Phi_{AC} = \Phi_{GA}, \qquad \Phi_{CG} = 0,    \label{SSConstraints}
\end{split}
\end{equation}
which reduces the number of independent parameters to six, as required of a strand-symmetric matrix~\cite{sueoka1995intrastrand}.  Most genomic sequences, when 
examined on a sufficiently large scale are observed to be strand-symmetric.  The parameter choices $\pi_a$ and $C_{ab}$ in the right-hand column of 
Table~\ref{tab:RateMatrixParams} are obtained by averaging pairs of parameters constrained to be equal by Eq.~(\ref{SSConstraints}).  
The one independent flux was then chosen arbitrarily within the range allowed by the constraint Eq.~(\ref{paramConstr}).  
The resulting rate matrix is 
\begin{equation}
Q^{\rm SS} = \left( \begin{array}{rrrr}
	-11.231 	& 2.154 		& 7.231 	& 1.846 \\
	1.143 	& -18.000 		& 0.571 	& 16.286 \\
	16.286 	&  0.571 		& -18.000	& 1.143 \\
	1.846 	&  7.231 		&  2.154 	& -11.231
\end{array} \right)    \times 10^{-4}.
\end{equation}
\item[SSR.] If the one remaining flux, namely $\Phi_{AC} = \Phi_{GA}$ which corresponds to a closed path $A \rightarrow C \rightarrow T \rightarrow G \rightarrow A$  
in Fig.~\ref{fig:Simplices}(b), is constrained to be zero, the resulting rate matrix is strand-symmetric and reversible.  
Such a matrix has five free parameters.  Setting  $\Phi_{AC} = \Phi_{GA} = 0$ 
while retaining the remaining parameters in the second column of Table~\ref{tab:RateMatrixParams} yields the strand-symmetric, reversible matrix 
\begin{equation}
Q^{\rm SSR} = \left( \begin{array}{rrrr}
	-11.231 	& 1.385 		& 8.000 	& 1.846 \\
	2.571 	& -18.000 		& 0.571 	& 14.857 \\
	14.857 	&  0.571 		& -18.000	& 2.571 \\
	1.846 	&  8.000 		&  1.385 	& -11.231
\end{array} \right)    \times 10^{-4}.
\end{equation}
\end{description}

\begin{table}[t]
\caption{Rate matrix parameters used in $K = 4$ numerical simulations.  The conventions of Eq.~(\ref{parameterList}) are used with the indices 1 to 4 representing 
the nucleotides $A$, $C$, $G$ and $T$ respectively.  The values of $\Phi_{ij}$ in the table refer to the non-reversible matrices $Q^{\rm GRM}$ 
and $Q^{\rm SS}$ only.  For the reversible matrices $Q^{\rm GTR}$ and $Q^{\rm SSR}$ all $\Phi_{ij}$ are zero.}
\begin{center}
\begin{tabular}{lrr}
			&	GRM and GTR		&	SS and SSR   \\
\hline
$\pi_A$		&    $0.400$				&     $0.325$		\\
$\pi_C$		&    $0.300$				&     $0.175$		\\
$\pi_G$		&    $0.050$			&     $0.175$			\\
\hline
$C_{AC}$		&    $1.5 \times 10^{-4}$ 	&     $0.9 \times 10^{-4}$	\\
$C_{AG}$		&    $1.6 \times 10^{-4}$ 	&     $5.2 \times 10^{-4}$	\\
$C_{AT}$		&    $1.2 \times 10^{-4}$ 	&     $1.2 \times 10^{-4}$	\\
$C_{CG}$		&    $0.2 \times 10^{-4}$ 	&     $0.2 \times 10^{-4}$	\\
$C_{CT}$		&    $8.8 \times 10^{-4}$ 	&     $5.2 \times 10^{-4}$	\\
$C_{GT}$		&    $0.3 \times 10^{-4}$ 	&     $0.9 \times 10^{-4}$	\\
\hline
$\Phi_{CG}$	&    $0.15 \times 10^{-4}$	&	$0$				\\
$\Phi_{GA}$	&   $-0.10 \times 10^{-4}$	&    $0.50 \times 10^{-4}$	\\
$\Phi_{AC}$	&    $1.00\times 10^{-4}$ 	&    $0.50 \times 10^{-4}$	\\
\hline
\end{tabular}
\end{center}
\label{tab:RateMatrixParams}
\end{table}%

\begin{figure}[t!]
   \centering
   \includegraphics[width=0.9\textwidth]{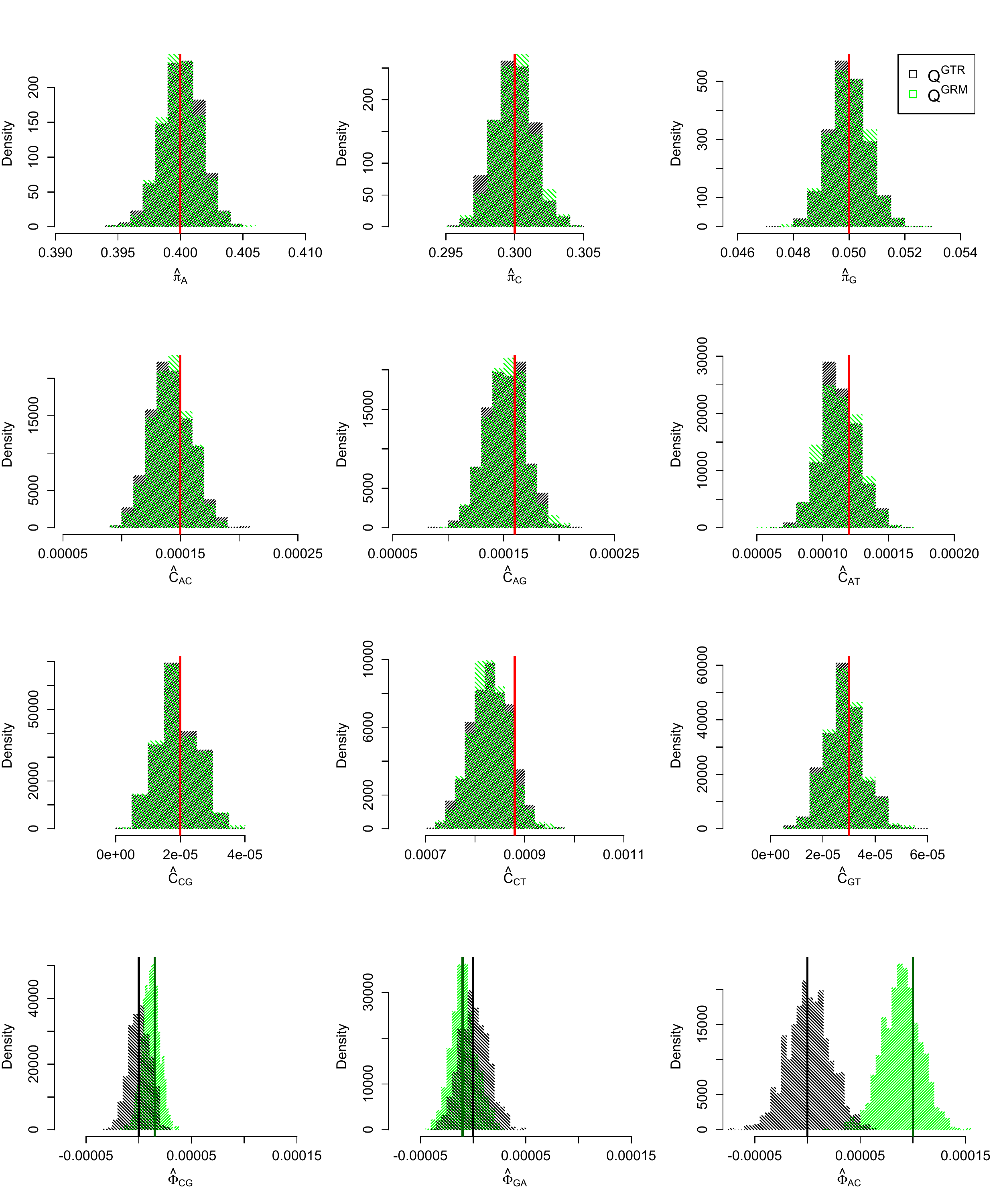} 
   \caption{Histograms of estimates of the parameters $\pi_i$, $C_{ab}$ and $\Phi_{ij}$ defining the rate matrices $Q^{\rm GTR}$ (black histograms) 
   and $Q^{\rm GRM}$ (green histograms) for $K = 4$ alleles.  True values of the parameters are given in the first column of Table~\ref{tab:RateMatrixParams} 
   and are indicated by the thick vertical lines.  
   1000 independent datasets were generated for each rate matrix assuming the data to be sampled from $M = 8$ independently chosen individuals from a 
   defining population of $N = 30$ individuals.}
   \label{ParameterEstimates4AllelesM8}
\end{figure}

\begin{figure}[t]
   \centering
   \includegraphics[width=\textwidth]{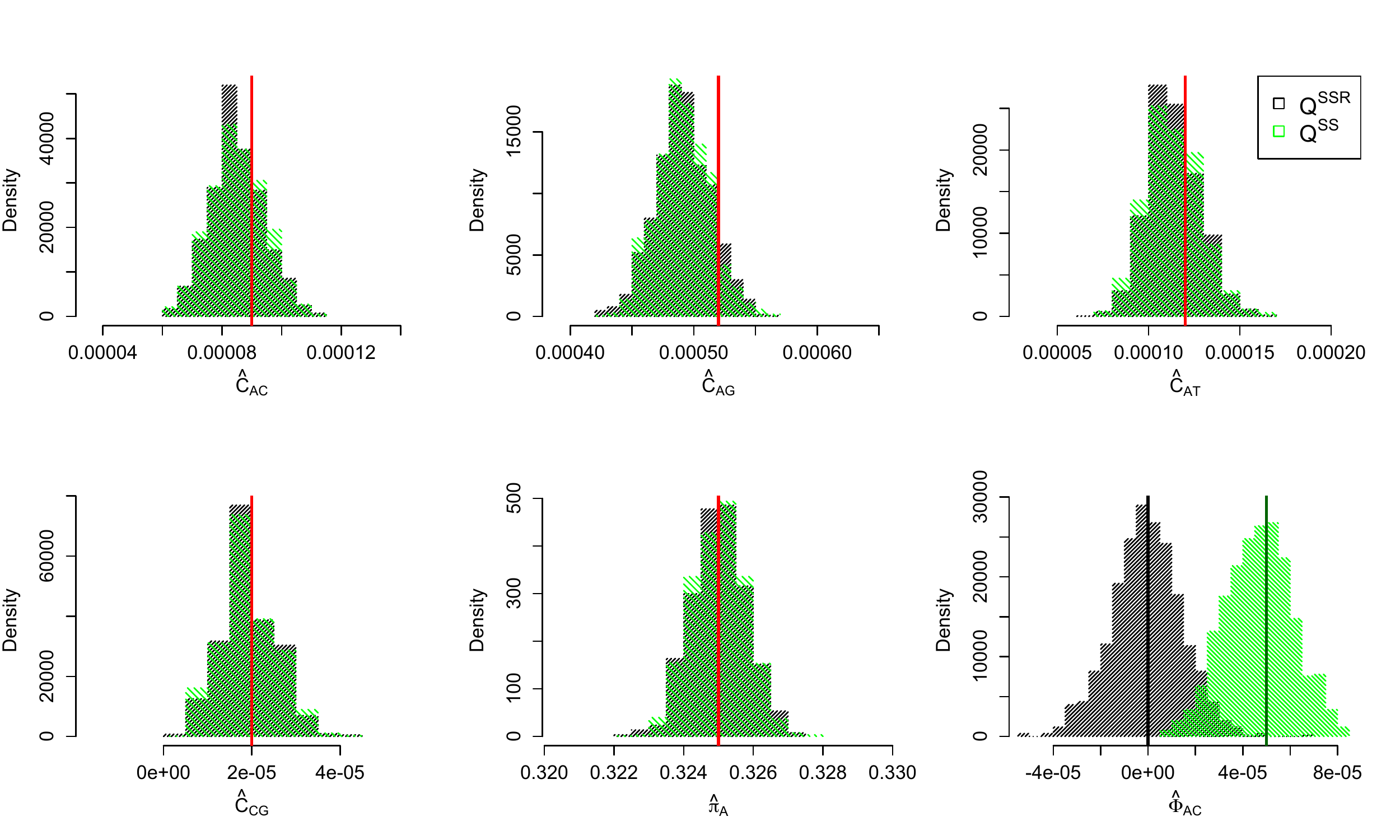} 
   \caption{Histograms of estimates of the parameters $\pi_A$, $C_{AC}$, $C_{AG}$, $C_{AT}$, $C_{CG}$ and $\Phi_{AC}$ defining 
   the strand-symmetric rate matrices $Q^{\rm SSR}$ (black histograms) and $Q^{\rm SS}$ (green histograms) for $K = 4$ alleles.  
   True values of the parameters are given in the second 
   column of Table~\ref{tab:RateMatrixParams} and are indicated by the thick vertical lines.  
   1000 independent datasets were generated for each rate matrix assuming the data to be sampled from $M = 8$ independently chosen individuals from a 
   defining population of $N = 30$ individuals.}
   \label{ParameterEstimates4AllelesSSM8}
\end{figure}

For each of the above four rate matrices a total of 1000 synthetic datasets each sampled at $10^5$ independent genomic sites were constructed using the 
procedure outlined for the $K = 3$ case in the previous section, 
except that the numerical experiment was carried out assuming a sample of $M = 8$ individuals.  
Maximum likelihood estimates of the parameters in the rate matrices $Q^{\rm GRM}$ and $Q^{\rm GTR}$ were calculated for each dataset using the theory of 
Section~\ref{sec:parameterEstimation}.  Maximum likelihood estimates of the parameters of the rate matrices $Q^{\rm SS}$ and $Q^{\rm SSR}$ were calculated 
using analogous likelihood functions constrained by Eqs.~(\ref{SSConstraints}).  Histograms of the estimated parameters are plotted in 
Figs.~\ref{ParameterEstimates4AllelesM8} and \ref{ParameterEstimates4AllelesSSM8}.  

\begin{figure}[h]
   \centering
   \includegraphics[width=\textwidth]{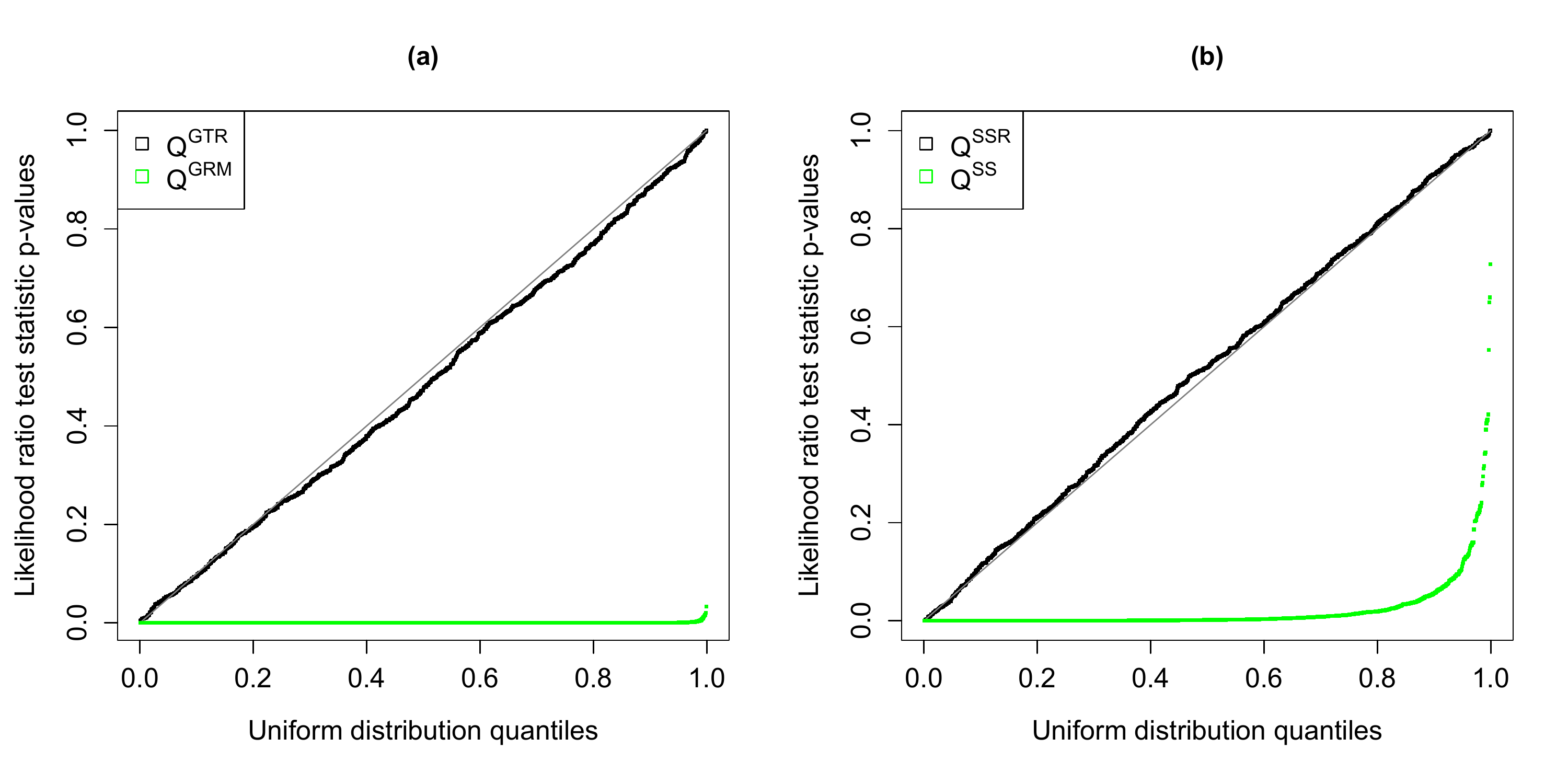} 
   \caption{QQ plots of 1-sided p-values calculated from the likelihood ratio statistic assuming a null hypothesis $\Phi_{ij} = 0$. (a) For the matrices $Q^{\rm GTR}$ 
   and $Q^{\rm GRM}$ likelihoods are maximised without contraints on any parameters, and p-values are calculated from the the likelihood ratio statistic assuming 
   a chi-squared distribution with 3 degrees of freedom.  (b) For the matrices $Q^{\rm SSR}$ and $Q^{\rm SS}$ likelihoods are maximised under the constraints of 
   Eq.~(\ref{SSConstraints}) and p-values are calculated assuming a chi-squared distribtuion with 1 degree of freedom.  The p-values are expected to have a 
   uniform distrbution for the reversible matrices $Q^{\rm GTR}$ and $Q^{\rm SSR}$, which satisfy the null hypothesis, but not for the non-reversible matrices 
   $Q^{\rm GRM}$ and $Q^{\rm SS}$.  }
   \label{fig:QQplots4AllelesM8}
\end{figure}

In common with the $K = 3$ case, estimates of the parameters $\pi_i$ and $C_{ab}$ defining the reversible part are independent of $\Phi_{ij}$, as expected.  
Also in common with the $K = 3$ case we observe that the $\hat\pi_i$ estimates are centred about their true value, but the estimates $\hat C_{ab}$ tend to be 
low.  A similar pattern was observed by Vogl and Bergman for $K = 2$ simulations with selection (see Fig.~3 of \cite{vogl2015inference}).  Estimates $\hat\phi_{ij}$ 
of the parameters determining the extent of non-reversibility are centred about their true values, or slightly skewed towards zero.  Figure~\ref{fig:QQplots4AllelesM8}
shows qq-plots of 1-sided p-values calculated from likelihood ratio test statistics analogous to Eq.~(\ref{KEq3LRTStatistic}) against a uniform distribution, assuming 
the null hypothesis $\Phi_{ij} = 0$.  For the otherwise unconstrained reversible and non-reversible rate matrices $Q^{\rm GTR}$ and $Q^{\rm GRM}$, the p-values were calculated 
assuming a chi-squared distribution with 3 degrees of freedom for the likelihood ratio test statistic.  For the strand-symmetric rate matrices $Q^{\rm SSR}$ and 
$Q^{\rm SS}$ a chi-squared distribution with 1 degree of freedom was assumed.  For the reversible matrices the p-values have a uniform distribution as required, 
while the p-values for the non-reversible rate matrices are suitably small.  


\section{Discussion and Conclusions}
\label{sec:Conclusions}

We have demonstrated that it is possible, in principle, to estimate an evolutionary rate matrix from the site frequency spectrum of an alignment 
of genomes sampled from a population, provided certain conditions are met.  The procedure consists of a maximum likelihood estimate of all 
12 parameters of the $4 \times 4$ nucleotide mutation rate matrix based on the distribution of observed single nucleotide polymorphisms at neutral sites in a multiple alignment.  
Given the site frequency spectrum constructed from the alignment of a large number of neutral sites obtained from sequencing a moderate number of individuals, 
the calculation is computationally straightforward and also provides a likelihood ratio test of the significance of the non-reversible part of the rate matrix.  

We feel it is important to visit in turn each of the conditions required of our procedure to highlight the remaining challenges inherent in the approach.  

Firstly we have assumed the population to have a constant size and to evolve according to the dynamics of a Wright-Fisher model.  Except in papers specifically 
addressing the point, this assumption is almost universally made implicitly throughout both the population dynamics and phylogenetics literature.  In practice, 
though, the assumption is often violated for real datasets.  It is important to recognise that a non-constant population size can alter the dynamics of 
genetic drift in ways which can effect the site frequency spectrum.    
For instance the recent explosive growth in the human population is believed to have skewed the site frequency spectrum 
towards rare numbers of derived alleles~\cite{Keinan11052012}.  On the other hand, if a rapidly growing population is modelled with Galton-Watson branching process, 
alleles present in the population at the beginning of the growth period can remain endemic in the population indefinitely without dissipating through genetic drift, 
even in the absence of new mutations at the spectrum boundary~\cite{burden2016genetic}.  We therefore caution that, as a general principle, the 
constant population Wright-Fisher model may be inappropriate for estimating mutation rates from populations of rapidly varying size.  

Secondly we have assumed that a large number $L$ of sites within a genome can be identified which have evolved independently and neutrally.  
Promising candidates are short intron sites~\cite{parsch2010utility} and 4-fold degenerate sites within codons, 
which are assumed to be relatively free of selective pressures.  Vogl and Bergman~\cite{vogl2015inference} have estimated selection effects in datasets consisting 
the short intron sites and 4-fold degenerate sites of a multiple alignment of 10 whole genome {\em Drosophila simulans} genomes.  
By partitioning the set of nucleotides into two effective alleles, $\{A, T\}$ and $\{C, G\}$, they are able to reduce the problem of estimating the parameters of 
a $K = 2$ Wright-Fisher model with both mutation and selection included, and find clear evidence of directional selection favouring the $\{C, G\}$ state.  
Therefore our analysis is not suitable for this particular {\em D.\ simulans} dataset, for instance.  
For consistency, any analysis of a real dataset to determine the full set of 12 parameters of a $4 \times 4$ mutation matrix would first have to pass Vogl and Bergman's 
test of neutrality.  

Thirdly we have assumed that the genomic sites considered evolve with a common rate matrix.  There is clear evidence however that biochemical effects 
can render mutation rates context dependent~\cite{zhao2002neighboring,siepel2004phylogenetic,hwang2004bayesian}, that is, 
rates may depend on the identity of neighbouring nucleotides.  
A strong example of this is the effect on $C \rightarrow T$ transitions of the $CpG$ context.  Assuming the context of a neutral site to be subject to selection 
pressures and therefore relatively stable, one could possibly accommodate context dependence by restricting the dataset, and hence the estimated rate matrix, 
to genomic sites with a particular context.  

Finally we mention the caveat that the estimation procedure relies on an approximate solution to the forward Kolmogorov equation valid in the limit of 
small scaled mutation rates, by which we essentially mean the Ewen-Watterson `$\theta$-parameter' of order $u_{ab} N$, where $N$ is the population size 
and $u_{ab}$ the per-generation mutation rate from allele $A_a$ to allele $A_b$.  As a rule of thumb, the approximate solution agrees well with numerical 
simulations of the neutral Wright-Fisher model provided $\theta < O(10^{-2})$~\cite{Burden:2016fk}.  


\bibliographystyle{elsarticle-num} 
\bibliography{PopulationGenetics}


\end{document}